# A Bioinspired Magnetothermally Triggered Capsule for Rapid Microrobot Release

*Mengfan Zhang[#], Zike Chen[#], Zhihao Lv, Zheng Jia, Rui Xiao*, and Guoyong Mao**

State Key Laboratory of Fluid Power and Mechatronic Systems, Key Laboratory of Soft Machines and Smart Devices of Zhejiang Province, Center for X-mechanics, Department of Engineering Mechanics, Zhejiang University, Hangzhou 310027, People's Republic of China

E-mail: rxiao@zju.edu.cn, guoyongmao@zju.edu.cn

**Abstract**

Capsules are attractive platforms for microrobot delivery, combining structural simplicity with protected delivery and rapid release remains challenging. Inspired by the latched spring mechanism of *Impatiens balsamina* seed dispersal, a magnetothermally triggered capsule integrates a phase-change hydrogel latch, a monostable cover that serves as an elastic spring, and a payload-carrying base. Mechanical analysis guides the cover design to avoid bistability after assembly and premature hydrogel latch rupture before triggering. During transport, the hydrogel constrains the inverted cover. Under an alternating magnetic field, $Fe_3O_4$-mediated heating softens the hydrogel and induces cohesive failure, releasing stored energy to drive capsule opening and microrobot ejection. The capsule opens within 50 ms after hydrogel softening and ejects microrobots at an estimated initial velocity of 1.0 m/s. Demonstrations on a three-dimensional platform and ex vivo porcine stomach establish protected delivery, target-site opening, and post-release actuation.

[#]Mengfan Zhang and Zike Chen contributed equally to this work.

## 1. Introduction

Microrobots have emerged as a versatile platform with increasingly sophisticated untethered actuation, programmable functionalities, and shape morphing capabilities,[1–4] enabling active interventions in confined biological environments, including targeted navigation, drug delivery, localized therapy, and in vivo imaging within the gastrointestinal (GI) tract.[5–8] Translating these capabilities into practical in vivo use requires not only functional microrobots but also reliable strategies for protected delivery and rapid release within the complex environment of the GI tract.[9,10] For example, microrobots must be shielded from acidic fluids, mucus adhesion, peristaltic folds, mechanical agitation, and premature dispersion.[11,12] Current delivery strategies primarily rely on conventional capsules, pills, hydrogels, or endoscopy-assisted delivery,[9–11,13] but achieving rapid release without added structural complexity remains challenging.

Many biological systems naturally combine protected delivery with rapid triggered release. For example, *Impatiens balsamina* employs an explosive seed dispersal mechanism **(Figure 1A)**. At maturity, a slight disturbance triggers rupture of the seed pod, releasing the stored tension and causing the valves to curl rapidly inward and eject the seeds, with related *Impatiens* species reaching speeds of approximately 1 m/s–4 m/s.[14,15] The valves both protect the seeds and store energy, while rupture simply releases the constraint, enabling the stored energy to drive rapid ejection without a complex mechanism. This process can be simplified as a latched spring model (Figure 1B),[16] in which the deformed valves act as an elastic spring that stores strain energy, while the surrounding tissue acts as a latch that maintains the deformed state before activation. When the latch is released, the stored elastic energy is rapidly converted into kinetic energy to drive seed ejection. A related mechanism has been demonstrated in a soft actuator, in which elastic energy stored in a light-responsive liquid crystal elastomer was released through a photothermally induced crystal-to-liquid transition of a liquid-crystalline adhesive latch, generating rapid launching motion.[17] This architecture translates naturally to capsule design: a closed capsule can protect and deliver its cargo through the GI tract, while stored elastic energy can drive rapid release and the trigger only releases the latch, reducing mechanical complexity. Existing capsule technologies have not fully exploited this architecture to combine protected delivery and rapid release within a simple, payload-adaptable design.[18–20] Instead, current systems generally follow two release paradigms: mechanical opening architectures and stimuli-responsive smart materials.

Mechanical capsules built with electronics, springs, gas reservoirs, and miniature beam actuators have been developed for controlled drug delivery.[21–23] Several systems can achieve on-demand release on the seconds timescale.[24,25] Magnetically actuated valves provide another approach for remotely controlled sampling and drug delivery in capsule systems.[26–28] However, the use of multiple miniaturized components in some systems, together with the predominant focus on drug delivery and sampling, may complicate their adaptation for microrobot loading and deployment. As an example, a gastric retentive robotic capsule integrates power, wireless communication, and multiple drug ejection modules within an 8.4 mm × 27 mm gelatin capsule shell, leaving limited space for microrobots.[24] In contrast, the use of stimuli-responsive smart materials enables a simpler design of the capsule-based carrier with fewer active components.[29,30] Without a dedicated structural amplifier, the resulting release is governed by the intrinsic kinetics of the material transformation and can unfold over minutes to hours.[31,32] To speed up the release rate, compressed spring-like structures are used,[33,34] as demonstrated by a compact mesoscale spring actuator. When paired with a barbed microneedle-anchored drug deposit, the actuator can propel the deposit into GI tissue within around 14 s.[34] For microrobot deployment, a key challenge is to integrate an elastic spring and a remotely triggered latch within a capsule for protected delivery and rapid release.

To address this challenge, we develop a magnetothermally triggered capsule inspired by the latched spring mechanism of *Impatiens balsamina* seed dispersal. The capsule couples a phase-change hydrogel latch with a monostable cover that acts as an elastic spring, enabling protected delivery and rapid on-demand ejection of microrobots without complex miniaturized mechanical components. Integrating these components introduces two mechanical requirements: the assembled cover must remain monostable to recover spontaneously after unlocking, and the hydrogel latch must withstand the restoring force of the inverted cover before triggering. First, we develop a gelatin-based magnetothermal hydrogel reinforced with carrageenan and magnetic nanoparticles to provide a stable mechanical constraint before rapid softening under an alternating magnetic field (AMF). Subsequently, finite element analysis (FEA) is used to guide the geometric design of the monostable soft cover, avoiding bistability and premature hydrogel rupture while enabling rapid elastic recovery upon unlocking. Finally, we demonstrate the integrated capsule for magnetic delivery, target site AMF-triggered opening, microrobot ejection, and magnetic deployment on both a 3D-printed gastrointestinal platform and ex vivo porcine stomach tissue. The triggered capsule opens within 50 ms after hydrogel softening and ejects microrobots at an estimated initial velocity of approximately 1.0 m/s. This design provides a simple and payload-adaptable framework for protected transport and rapid on-demand deployment of microrobots and other functional payloads in confined biological environments.

## 2. Results and Discussion

### 2.1. Design of the capsule

To translate the latched spring mechanism into a functional trigger-release process, the capsule is designed with three core components (Figure 1C and D): a monostable soft cover, a payload-carrying 3D-printed polylactic acid (PLA) base containing microrobots, and a magnetothermally triggered phase-change hydrogel layer. The soft cover functions as the elastic energy reservoir, the PLA base provides a protected chamber for microrobots, and the hydrogel layer acts as a remotely triggered latch that controls the timing of release.

During assembly, the soft cover is mechanically inverted and bonded to the hydrogel layer anchored on the PLA base. This inversion stores elastic strain energy in the cover, while the hydrogel layer maintains the cover in the inverted state before triggering. After the capsule reaches the target site, an external AMF is applied to release the hydrogel constraint. $Fe_3O_4$ nanoparticles ($Fe_3O_4$ NPs) in the hydrogel convert electromagnetic energy into localized heat, inducing hydrogel softening. As the hydrogel loses its mechanical integrity, cohesive failure occurs within the hydrogel layer and releases the inverted cover. The monostable cover then spontaneously recovers to its original configuration, converting the stored elastic strain energy into kinetic energy that drives rapid capsule opening and microrobot ejection.

The intended workflow for protected delivery and rapid on-demand deployment in the GI tract is illustrated in Figure 1E. After oral administration, the closed capsule travels into the stomach and can be magnetically guided along the gastric surface toward the target region while preventing premature exposure or dispersion of the microrobots. Magnetic guidance and magnetothermal triggering were performed sequentially using separate field sources: a quasi-static magnetic field for navigation and the AMF for $Fe_3O_4$ NPs-mediated heating. Once the capsule reaches the target site, AMF triggers hydrogel softening and capsule opening. The released magnetic microrobots can then be further guided by external magnetic actuation for localized treatment.

Considering the thermal, safety, and biocompatibility requirements of the capsule, gelatin was selected as the hydrogel matrix because of its favorable biocompatibility and thermally induced gel-sol transition. However, pure gelatin hydrogels have limited mechanical strength[35] and relatively low transition temperatures (30 °C–36 °C). Accordingly, the capsule design involved two coupled requirements. First, the hydrogel layer needed to remain mechanically stable at physiological body temperature while undergoing rapid softening under AMF triggered heating. Second, the soft cover needed to exhibit facile elastic recoverability and reliable hydrogel-mediated immobilization. These material and structural requirements are addressed in the following sections.

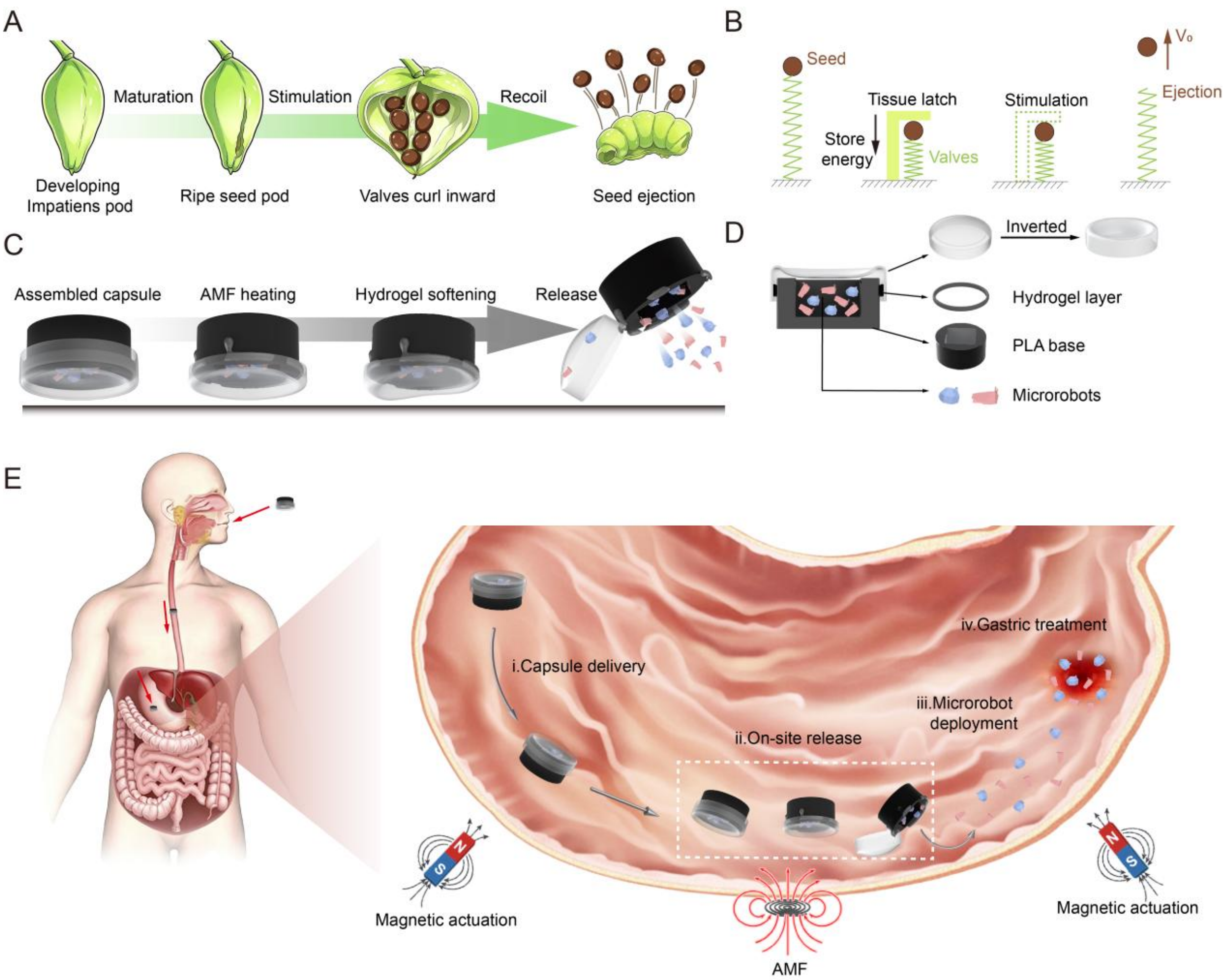

**Figure 1. Design concept and working principle of a magnetothermally triggered capsule inspired by the explosive seed dispersal of *Impatiens balsamina*.** (A) Schematic illustration of explosive seed dispersal in *Impatiens balsamina*. After maturation, mechanical stimulation triggers rapid valve curling in the ripe seed pod, leading to seed ejection. (B) Simplified latched spring model of the seed ejection mechanism. (C) Schematic illustration of AMF-triggered hydrogel softening that releases the inverted cover for capsule opening and microrobot ejection. (D) Exploded view of the capsule structure. The capsule consists of a monostable soft cover, a magnetothermally triggered phase-change hydrogel layer, a payload-carrying base housing the cargo, and the magnetic microrobots. (E) Proposed gastric operation workflow, including protected capsule delivery, on-site AMF-triggered release, microrobot deployment, and localized treatment.

### 2.2. Magnetothermally triggered hydrogel

The hydrogel layer must remain mechanically robust at 37 °C to constrain the inverted cover during transit, yet soften or even liquefy rapidly upon AMF-triggered heating to release the cover. For this, we designed a gelatin-based magnetothermally triggered phase-change hydrogel containing carrageenan and $Fe_3O_4$ NPs, referred to as the GC-$Fe_3O_4$ hydrogel (**Figure 2A**). At the molecular level (Figure 2B), carrageenan can interact with gelatin through noncovalent interactions, particularly electrostatic interactions between the negatively charged sulfate groups of carrageenan and positively charged groups of gelatin, as well as hydrogen bonding.[36,37] These interactions promote conformational ordering of gelatin and carrageenan chains and introduce additional junction zones within the protein-polysaccharide gel network,[38] thereby improving the mechanical integrity of the gel state and increasing the temperature required for network dissociation. In parallel, $Fe_3O_4$ NPs act as the magnetic heating component,[39] converting AMF input into localized heat to drive hydrogel softening on demand.

To examine the structural basis of this mechanism, we characterized the hydrogel network using FTIR and XRD analyses. FTIR spectra confirmed the presence of characteristic peaks for gelatin, carrageenan, and $Fe_3O_4$ NPs in the GC-$Fe_3O_4$ hydrogel, including the characteristic amide bands (3438 $cm^{-1}$, N-H and O-H, 1642 $cm^{-1}$, C=O) of gelatin, the sulfate-related S=O vibration (1237 $cm^{-1}$) of carrageenan, and the Fe-O vibration (602 $cm^{-1}$) of $Fe_3O_4$ NPs (Figure 2C), indicating the successful compositing of the three components. Compared with pure gelatin (3445 $cm^{-1}$, N-H and O-H, 1644 $cm^{-1}$, C=O), the Amide A and Amide I bands of GC-$Fe_3O_4$ hydrogel (3438 $cm^{-1}$, N-H and O-H, 1642 $cm^{-1}$, C=O) shifted to a lower wavenumber, consistent with strengthened electrostatic interactions and hydrogen bonding in the composite network.[38] The enhanced vibrational intensity of the Amide I band in GC-$Fe_3O_4$ hydrogel further suggests changes in the protein-associated secondary structure, which may be associated with a more ordered helical structure in hydrogels.[40] XRD patterns (Figure 2D) showed that the GC-$Fe_3O_4$ hydrogel retained the characteristic diffraction peaks of $Fe_3O_4$ NPs, confirming successful nanoparticle incorporation.[41] Meanwhile, the crystalline peaks of carrageenan were largely suppressed in the composite hydrogel, and the intensity of the broad diffraction peaks associated with the gelatin also decreased. These changes indicate reduced long-range crystalline order of gelatin and carrageenan after composite formation, consistent with enhanced local intermolecular associations within a less crystalline composite network.

We further characterized the hydrogel from two aspects for remote constraint release: its intrinsic thermally triggered liquefaction behavior and its AMF-triggered heating capability. The former determines the temperature range at which the hydrogel loses its gel-state mechanical integrity, whereas the latter determines whether $Fe_3O_4$ NPs-mediated magnetothermal heating can raise the hydrogel to this transition range during capsule operation. The developed hydrogels with different compositions are listed in Table S1. As shown in Figure 2E, the thermally triggered liquefaction behavior of the hydrogels was characterized by temperature-dependent rheological analysis. The pure gelatin hydrogel showed a rapid decrease in storage modulus (G') from 30 °C to 36 °C, indicating loss of the gel network at relatively low temperature. Incorporation of $Fe_3O_4$ NPs alone did not provide sufficient thermal stability, as the gelatin-$Fe_3O_4$ hydrogel, referred to as the G-$Fe_3O_4$ hydrogel, exhibited a similarly rapid decrease in G'. In contrast, the GC-$Fe_3O_4$ hydrogel maintained a higher G' over a broader temperature range, and its modulus exhibited a step-like decrease, remaining at approximately 1 kPa at 37 °C. This indicated that carrageenan reinforced the gelatin network and increased the temperature required for gel-sol transition.[36] The further increase in carrageenan content shifted the gel-sol transition to a higher temperature, confirming the tunability of the hydrogel thermal response. These results indicate that carrageenan improves the mechanical integrity and the temperature required for network dissociation of the hydrogel layer under physiological conditions while preserving heat-triggered softening for capsule opening.

As shown in Figure 2F, the magnetic hysteresis curve confirmed that $Fe_3O_4$ NPs remained magnetically active after incorporation into the hydrogel network. Although the GC-$Fe_3O_4$ hydrogel showed finite saturation magnetization (14.7 $emu \cdot g^{-1}$) and coercive field (125.7 Oe), its saturation magnetization was sufficient to support magnetic responsiveness and AMF-triggered heating. Furthermore, the hydrogels with varying concentrations of $Fe_3O_4$ NPs were formed into 1-mm-thick discs, and their temperature changes under AMF were recorded. From an initial temperature of 21.0 °C, the samples with varying concentrations of $Fe_3O_4$ NPs, subjected to AMF for 60 s, reached temperature ranging from 30.8 °C to 41.4 °C, while the pure gelatin hydrogel showed negligible heating (Figure 2G, H). To further improve local heat accumulation, a thin silicone oil layer was applied to the hydrogel surface. Owing to its hydrophobicity and low thermal exchange with the surrounding aqueous environment, this coating reduced heat dissipation from the hydrogel and accelerated the temperature rise. Correspondingly, the temperatures of GC-$Fe_3O_4$ hydrogels under AMF for 60 s increased to higher ones from 41.1 °C to 56.1 °C. These results show that the higher amount of $Fe_3O_4$ NPs incorporated into the hydrogels brings enhanced AMF-triggered heating capability. Also, the silicone oil coating acts as an interfacial thermal barrier that concentrates AMF-triggered heat within the hydrogel, accelerating the gel-sol transition.

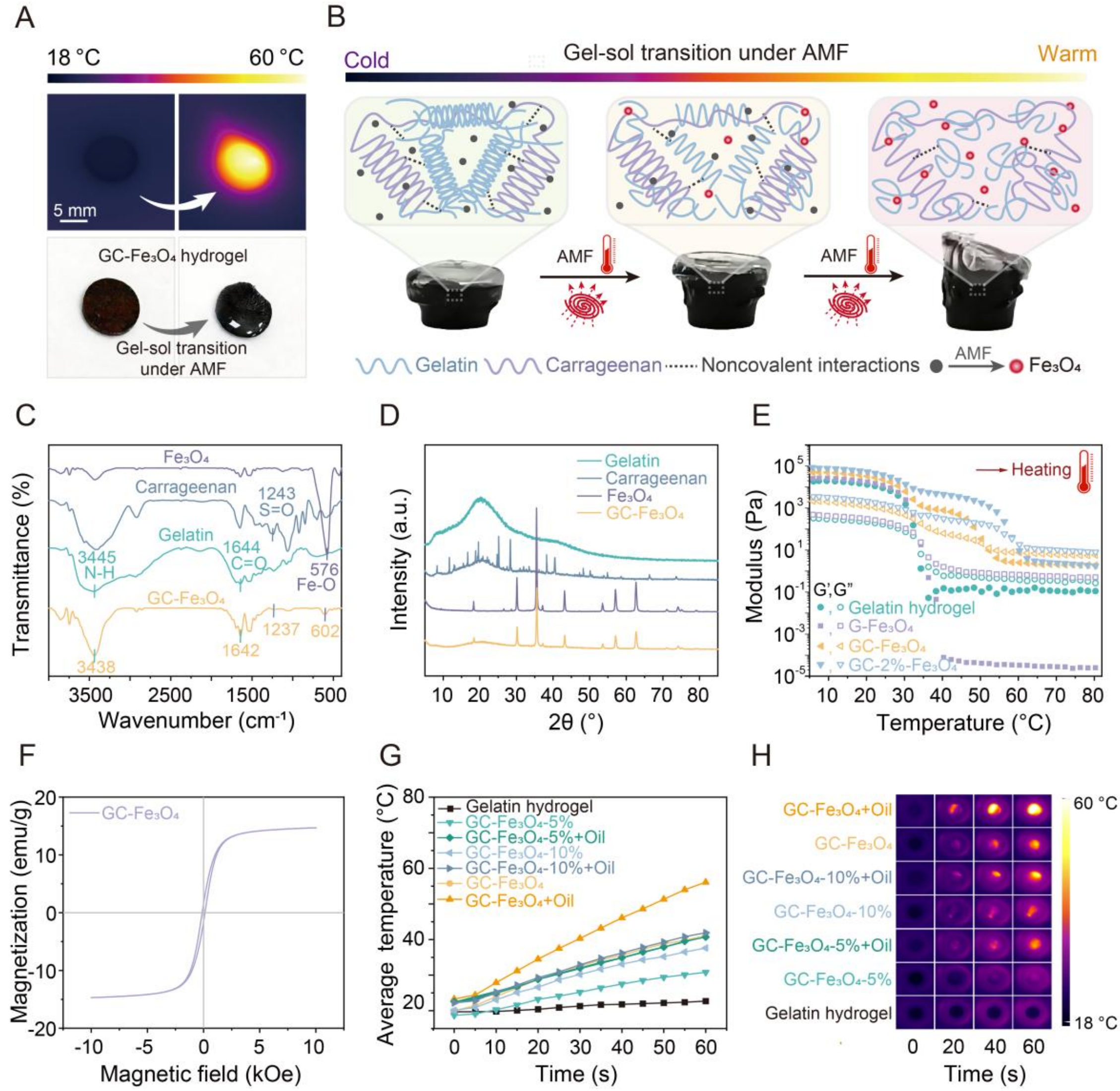

**Figure 2. Characterization of the magnetothermally triggered hydrogels for remote constraint release.** (A) Optical and infrared thermal images of GC-$Fe_3O_4$ hydrogel after 1 min of AMF exposure, showing hydrogel softening. (B) The molecular-chain mechanism underlying this softening behavior of GC-$Fe_3O_4$ hydrogel for capsule opening. (C) FTIR spectra. (D) XRD spectra. (E) The storage and loss moduli of the hydrogels as functions of temperature. Filled and open symbols represent the storage modulus and loss modulus, respectively. (F) The magnetic hysteresis curve of GC-$Fe_3O_4$ hydrogel. (G) AMF-triggered heating curves of hydrogels with different $Fe_3O_4$ concentrations and silicone oil coating condition. (H) Corresponding infrared thermal images.

### 2.3. Design of the monostable soft covers

To endow the cover with reliable packaging and cargo releasing capabilities, the cover must remain constrained by the hydrogel layer during transport and recover rapidly once the hydrogel is softened by heating. We harness the stability characteristics of soft covers to drive this releasing response. Unlike a bistable cover that requires an additional stimulus to switch between stable configurations,[42] a monostable cover stores elastic strain energy upon inversion and spontaneously reverts to its initial configuration once the physical constraint is removed. Therefore, the monostable cover is imperative for our design. Although the bifurcation and stability of idealized circular shells have been extensively studied,[43–45] translating these classical rules into our covers for applications introduces two coupled fundamental challenges. First, practical assembly necessitates a bonder structure (**Figure 3A**). The presence of this bonder significantly alters the boundary conditions, and its influence on the stability of the cover needs to be evaluated. Second, the inherent restoring strain energy of the inverted cover poses a severe risk of prematurely fracturing the hydrogel layer prior to thermal actuation.

Because the capsule size is constrained by biomedical application requirements, the soft cover must maintain a finite thickness to provide mechanical robustness, protect the internal cargo, and store sufficient elastic energy during inversion. Therefore, its geometry cannot be simplified as an ideal thin-shell structure. The cover comprises a main dome (the half span $L$, the height $H$, and the thickness $t$) and the bonder for assembly (Figure 3A). We propose three design strategies for the cover design. A direct extension of the thick dome in Design 1 generates an excessive strain energy in the hydrogel layer during the assembly of the capsule that inevitably tears the hydrogel layer, which is quantified by a subsequent finite element analysis. To reduce the energy after inversion, we considered two designs by reducing the thickness of the bonder structures. While shifting the centers of the inner and outer arcs (Design 2) thins the bonder, it undesirably couples the bonder dimensions with the main dome geometry. Therefore, we employ a decoupled geometric strategy (Design 3): by selectively extending only a portion of the dome edge, we independently program the bonder's thickness $T_b$ and length $L_b$ without altering the dome's intrinsic mechanics. This geometric decoupling is crucial for programming the system's stability (Figure 3B, C). A bistable soft cover passively locks into the inverted state as shown in Figure 3B and Movie S1. It would entrap the cargo even after liquefaction of the hydrogel. Conversely, the prescribed monostable soft cover converts stored strain energy into kinetic energy, guaranteeing spontaneous deployment (Movie S1). High-speed kinematic tracking (Figure 3C and Movie S2) demonstrates that, upon the removal of the boundary constraint, imposed by tweezers in this test, the cover undergoes rapid recovery, and is propelled away from the base with an average recovery speed of 1.21 m/s (Figure S1), completely detaching from the base within 100 ms.

We next performed FEA to reveal the underlying deformation mechanisms. To capture the limit point buckling and complex post buckling behaviors, the model was solved in Abaqus using the modified Riks method, with a rotational displacement applied to the bottom edge of bonder. The fundamental distinction between the bistable soft cover and monostable soft cover is governed by their strain energy ($U_s$) characteristics, evaluated using a dimensionless energy parameter $U_s^* = U_s/EL^3$, where $E$ is the Young's modulus of PDMS. The bistable soft cover is characterized by a non-monotonic energy curve featuring a local energy minimum (Figure 3D). Once the applied load drives the system over the critical snap-through point, it becomes physically trapped in this secondary energy well upon unloading, passively locking the inverted configuration.[46] As for monostable soft cover, the strain energy increases with the external load monotonically and will revert to the initial configuration after releasing, due to the monotonicity of the energy curve (Figure 3E).

To systematically elucidate how geometric parameters govern these stability mechanisms, we mapped the phase space of the decoupled structure (Figure 3F). To isolate the morphological effects of the main dome, we constrained the half span of soft covers to the predefined capsule base ($L$ = 5 mm) and imposed a baseline bonder with fully extended edge (identical to Design 1, with $L_b$ = 2 mm). Sweeping the normalized height ($H/L$) and thickness ($t/L$) under these conditions revealed that the added boundary thickness fundamentally shifts the phase boundary toward the bistable regime compared to a classical circular shell. Specifically, with the addition of the bonder part, the critical boundary shifted upward from the black dashed line $t/L = 0.22H/L - 0.002$ (fitted according to Madhukar *et al.*[43]) to the green dashed line $t/L = 0.5H/L+0.03$. The deviation of the two critical lines demonstrates that, with the appearance of bonder structure, the stability of the system tends to be bistable, meaning that the thickness $t$ needs to be greater than that of traditional circular shells to possess a monostable stability. Closely around this critical boundary appeared a pseudo-bistable phase, a topological anomaly where the energy curve exhibits a stationary point of inflection (Figure S2), causing the structure to revert instantaneously under small perturbations. Furthermore, configurations exceeding the upper boundary ($t/L = H/L$) fall into a geometrically impossible zone, which can be deduced from the design drawing in Figure 3A (Design 3). Since the monostable stability is the target mechanism, and a thick dome may result in a large resistance force, leading to rupture of hydrogel layer, we anchored the main part parameters close to the critical line as $H$ = 2.6 mm, $t$ = 1.5 mm within the monostable regime. Fixing this optimal dome geometry, we then expanded the analysis to the independent bonder parameters (Figure 3G). This second phase map unveils a bonder induced instability boundary defined by $T_b = -0.1L_b+1.15$. Emerging beyond a critical length threshold of $L_b^c$ = 2.5 mm, traversing this boundary implies that the local boundary stiffness dominates the global dome elasticity, leading to bistable instability. A geometrically impossible region also exists because $T_b$ must not be larger than natural length of the bottom edge, predefined by main part. Validated by experimental prototypes, these deterministic phase maps provide a rational design basis and reduce empirical trial and error. In practice, reliable assembly of the monostable cover and base requires a sufficiently long bonder. Otherwise, fabrication becomes impractical. Therefore, in the next section, the second problem will be discussed based on the same column where $L_b$ = 2.0 mm. This column presents a full column of monostable structure with a relatively long bonder.

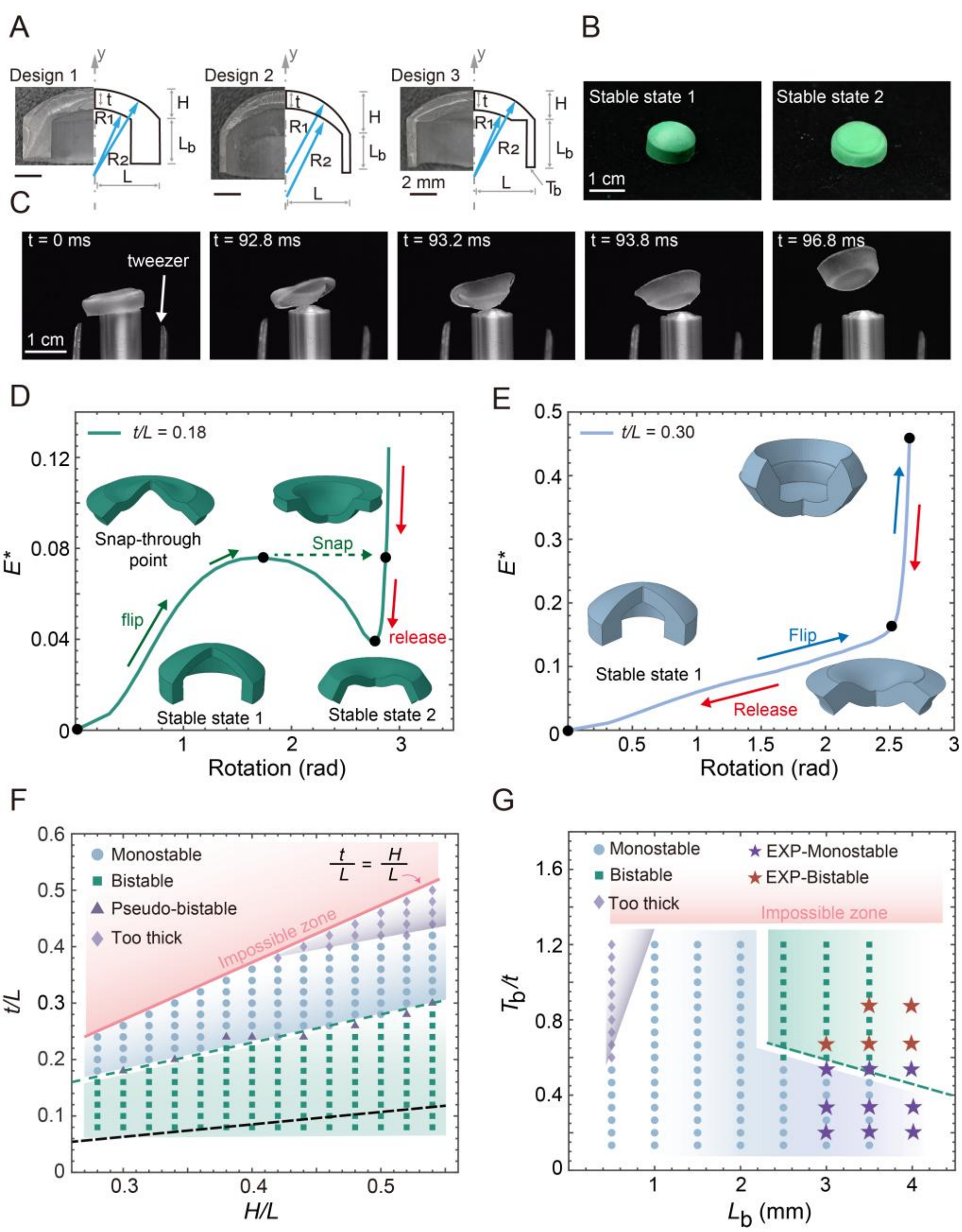

**Figure 3.** Geometric programming and stability phase map for soft covers. (A) Three geometric designs of soft covers. (B) Optical images of a representative bistable cover under its two stable

states. (C) Snapshots of a representative monostable cover with spontaneous recovery after release. (D) Strain energy curve of bistable cover with a deep energy well depicting the inversion process. (E) Monotonic strain energy curve depicting the inversion and spontaneous cover recovery process. (F) Phase map of transition between bistable and monostable regions in the normalized geometric parameters space of the main part. (G) Phase map of transition between bistable and monostable regions in the normalized geometric parameters space of bonder, along with data points (Star marks) indicating experimentally tested samples and their observed post-inversion behaviors. The corresponding cover parameters are provided in Table S2.

### 2.4. Geometric optimization of the soft cover for robust bonding

We next addressed the second fundamental challenge: matching the restoring energy of the inverted monostable soft cover with the fracture resistance of the hydrogel layer to prevent premature capsule opening at room temperature. Because the inverted cover stores elastic strain energy after assembly, it continuously applies a restoring driving force to the hydrogel layer before AMF triggering. Untriggered opening may occur through either interfacial debonding or cohesive fracture within the hydrogel. To resolve this, we first tailored the material properties of the hydrogel layer. While incorporating $Fe_3O_4$ NPs into the pure gelatin hydrogel (G-$Fe_3O_4$ hydrogel) successfully imparted necessary magnetothermal capabilities, it concurrently altered the mechanical properties and significantly stiffened the matrix (Young's modulus increased from 20.2 kPa to 84.5 kPa, **Figure 4A**). To reinforce the gelatin network and regulate its thermal transition, carrageenan was introduced into the polymer network to form the GC-$Fe_3O_4$ hydrogel. This integration further increased the Young's modulus from 84.5 kPa to 106.1 kPa and provided a stronger material matrix for subsequent structural coupling. Scanning electron microscopy (SEM) observations further supported the formation of a denser interconnected porous structure with $Fe_3O_4$ NPs embedded in the matrix, consistent with its increased Young's modulus (Figure S3).

To identify the dominant failure mode of the hydrogel layer and quantify its fracture resistance, we performed 180° peel tests at room temperature (Figure 4B). Specifically, two sets of peel tests were conducted to evaluate the effects of hydrogel composition (Figure 4C, E) and hydrogel thickness (Figure 4D, F). For all samples, the peel force was normalized by the hydrogel width $w_h = 25$ mm to minimize the influence of sample geometry. During peeling, the normalized force first reached a peak associated with crack initiation and then decreased to a steady plateau during stable crack propagation. The crack propagated within the hydrogel matrix rather than along the interface (Figure 4B and Movie S3), indicating that the interfacial adhesion was stronger than the cohesive strength of the hydrogel. Therefore, the measured peel response reflects cohesive fracture within the hydrogel layer rather than interfacial debonding.

Based on this cohesive fracture mode, the critical energy release rate $G_c$ and the initiation stress $\sigma_f$ were calculated from the steady-state peel response using the method described in Text S1.4. Here, $G_c$ characterizes the energy required to propagate an existing crack through the hydrogel matrix, whereas rupture of an intact hydrogel layer before triggering should be evaluated using the work to rupture,[47] $W^*$, or the maximum principal tensile stress criterion.[48] Failure occurs when the strain energy density or maximum principal tensile stress reaches the threshold $W^*$ or maximum tensile stress $\sigma_m$ before rupture, respectively. $W^*$ was calculated from the area under the nominal stress-stretch curve in uniaxial tensile rupture tests, equivalently the mechanical work to rupture divided by the initial gauge volume. $\sigma_m$ is defined as the maximum tensile stress. $W^*$ was used as the primary energy density threshold for assessing premature rupture of the intact hydrogel in FEA, while $\sigma_m$ provided a stress based qualitative check. $G_c$ was further used to estimate $\sigma_f$ as an auxiliary indicator for local damage initiation, rather than as a direct fracture threshold for intact material rupture.[49]

As shown in Figure 4A, 4C and 4E, adding $Fe_3O_4$ NPs and carrageenan sequentially increased the $\sigma_m$, $\sigma_f$ and $G_c$. More importantly, the GC-$Fe_3O_4$ hydrogel showed a substantially higher $W^*$ (0.034 mJ/mm$^3$) than the G-$Fe_3O_4$ hydrogel (0.025 mJ/mm$^3$) and pure gelatin hydrogel (0.0037 mJ/mm$^3$), making it the optimal hydrogel layer candidate. The displacement at the peak peel force decreased after introducing $Fe_3O_4$ NPs and carrageenan, which can be interpreted using the flaw sensitivity length,[47] defined as $\frac{G_c}{W^*}$. The flaw sensitivities of three materials were computed and listed in Table S3. Because the three hydrogels had comparable $G_c$ values but markedly different $W^*$ values, pure gelatin had a larger flaw sensitivity length and could tolerate a larger peeling induced flaw before stable crack growth, causing the peak force to appear at a larger displacement. Varying the hydrogel thickness changed the peak displacement and apparent $\sigma_f$ because of localized geometric effects, but $G$c remained nearly constant (Figure 4D, F). Although the 2 mm-thick GC-$Fe_3O_4$ hydrogel produced a higher $\sigma_f$, excessive thickness compromises gel-sol transition kinetics and capsule compactness. Balancing these competing factors, a 1 mm-thick GC-$Fe_3O_4$ hydrogel layer was selected as the optimal hydrogel layer. With the optimized hydrogel layer as the rupture resistance baseline, we next used FEA to guide the geometric design of the cover. Specifically, the simulations quantified the strain energy density imposed on the hydrogel after assembly with the inverted cover and evaluated whether this stored elastic energy could cause cohesive rupture before AMF triggering. The cover was first mechanically inverted with the hydrogel tied to the cover bonder, after which a stress-release step was simulated by fixing the outer edge of the hydrogel to represent adhesive attachment to the capsule base (Figure 4G, i-iii). Because the peel tests indicated that cohesive failure was more likely to occur within the hydrogel matrix, the contour plot in Figure 4G(iv) focuses on the strain energy density distribution from the central region toward the bonder side. The strain energy density was further extracted along the mid-plane path of the hydrogel, defined as the dangerous section for quantitative comparison.

Among the three cover designs (Figure 3A), Design 1 induced a strain energy density more than one order of magnitude higher than the other configurations and far exceeded $W^*$, validating its exclusion (Figure 4H, I). By contrast, Designs 2 and 3 both maintained energy density below $W^*$. Although Design 2 gave a slightly lower energy level, Design 3 was selected because its bonder parameters could be tuned independently. The parameters of covers with different designs are listed in Table S4. The strain energy density of Design 3 increased with bonder thickness $T_b$, surpassing the rupture threshold when $T_b$ exceeds 0.9 mm (Figure 4J). The maximum principal tensile stress in Figure S4 showed the same trend: Design 1 exceeded the maximum tensile stress $\sigma_m$, whereas Designs 2 and 3 remained below $\sigma_m$ denoting a safe bonding of hydrogel. In Design 3, the local stress was slightly higher than $\sigma_f$, suggesting that limited damage initiation may occur, but the hydrogel would remain intact after assembly. This behavior is acceptable because the hydrogel functions as a sacrificial latch that is designed to lose mechanical integrity during actuation. The same trend of stress is observed for different values of $T_b$ in Figure S4B, with hydrogel rupture predicted when $T_b$ exceeds 0.9 mm. It should be noted that the maximum principal tensile stress criterion is more suitable for brittle failure than for ductile materials.[48] Thus, this stress-based comparison is used only as a qualitative assessment and is less precise than $W^*$. Conversely, increasing the bonder length $L_b$ further reduced the strain energy density (Figure S5, simulated based on Design 3 with other parameters of $L = 5$ mm, $H = 2.6$ mm, $t = 1.5$ mm, $T_b = 0.8$ mm ), indicating improved adhesion stability. Combining these computational results with manufacturing feasibility, the final cover parameters were set as $L = 5$ mm, $H = 2.6$ mm, $t = 1.5$ mm, $L_b = 3.5$ mm, $T_b = 0.8$ mm, which were used in the experimental demonstration in the next section. This configuration prevented premature hydrogel rupture and enabled robust capsule assembly, as confirmed by fabricated prototypes.

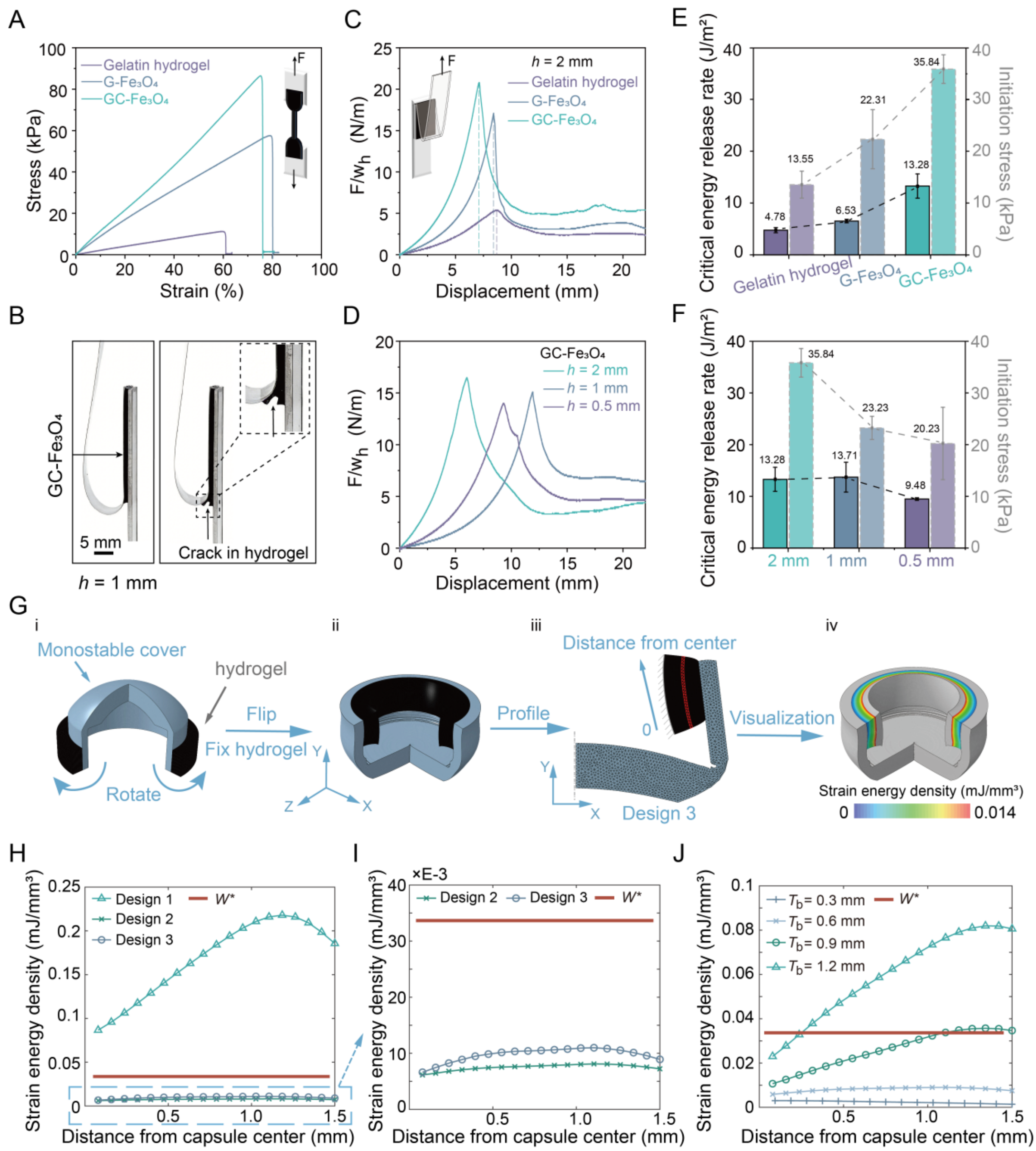

**Figure 4. Mechanical matching between work to rupture ($W^*$) of hydrogel and soft cover restoring energy density along the dangerous section through material characterization and FEA.** (A) Stress-strain curves of hydrogels with different material composition. (B) Optical images of the 180° peel test for measuring cohesive fracture of the hydrogel. (C) Normalized peel force-displacement curves of hydrogels with different material compositions. (D) Normalized peel force-displacement curves of GC-$Fe_3O_4$ hydrogels with different thicknesses. (E) Critical energy release rate and initiation stress of hydrogels with different material compositions. (F) Critical energy release rate and initiation stress of GC-$Fe_3O_4$ hydrogels with different thicknesses. (G) Finite element analysis process for strain energy density in hydrogel:

i) Assembly of cover and hydrogel in the initial step. ii) Simulated configuration of assembled cover after flipping, which is used as initial configuration for next step. iii) Boundary settings of flipped cover shown in the profile section, the outer edge of the hydrogel is set as fixed to simulate the attachment to capsule. The strain energy density data along the path in the middle of hydrogel are extracted for further rupture analysis. iv) Contour plot for strain energy density in hydrogel, which starts from middle (dangerous section) to the bonder side. (H, I) Comparison of strain energy density along middle path of hydrogel for three different cover designs. (J) Comparison of strain energy density curves along middle path of hydrogel for Design 3 with different values of $T_b$.

**2.5. Release and deployment of microrobots**

Having validated the key material and structural mechanisms, including magnetothermal constraint release and monostable cover recovery, we next evaluated the integrated capsule in system-level deployment of microrobots. A high-speed camera was first used to capture the AMF-triggered release process (**Figure 5A** and Movie S4). Before triggering, the capsule remained closed, with the inverted cover constrained by the hydrogel layer. Under the AMF, localized magnetothermal heating softened the hydrogel layer and weakened its constraint on the inverted cover. The resulting cohesive failure propagated through the hydrogel layer, leading the inverted cover to rapidly recover toward its initial configuration, enabling capsule opening within 50 ms after hydrogel softening (Figure 5A). Importantly, this sudden release of elastic strain energy generated a flipping motion of the entire capsule, which provided an ejection impulse for the microrobots while simultaneously creating a large release aperture. A similar AMF-triggered opening behavior was also observed for the capsule in the upright configuration (Figure S6A and Movie S5), where cover flipping produced a large opening without immediate microrobot ejection.

We then demonstrated the feasibility of the capsule for microrobot delivery, remote on-demand release, and post-release deployment on a 3D-printed platform. The operation consisted of three sequential steps (Figure 5B–D and Movie S6). First, the capsule was guided to the target location by a static magnet while maintaining its closed state. Because the cover was positioned on the lower side of the capsule, its soft surface reduced friction during delivery, allowing the capsule to reach the target site within approximately 70 s (Figure 5B). Second, the AMF was applied to heat the hydrogel layer, increasing the local temperature from approximately 20 °C to 41 °C (Figure 5C), causing capsule opening. The released elastic strain rapidly ejected the microrobots together with the PLA base. Because the microrobots remained with the PLA base during the initial ejection stage, their initial ejection velocity was estimated from the translational motion of the PLA base, yielding a fitted value of 1.017 m/s (Figure S7). This estimated microrobot ejection velocity is comparable to the lower end of the reported seed ejection velocities of *Impatiens* species (approximately 1 m/s–4 m/s),[14,15] indicating that the capsule reproduces the rapid energy release behavior of the biological system on a similar velocity scale. Upon landing, the PLA base impacted the platform, causing the microrobots to separate and disperse around the deployment site. Finally, the exposed microrobots, represented here by NdFeB magnetic elastomer pieces, were further guided by the static magnet to designated positions (Figure 5D). These results demonstrate that the capsule can integrate magnetic delivery, remote magnetothermal triggering, cargo ejection, and post-release magnetic control into a continuous deployment process. We further evaluated the capsule in an upright configuration (Figure S6B and Movie S7). In this orientation, the cover similarly opened after constraint release, but the microrobots remained within the exposed payload chamber and were subsequently extracted and guided out by the external static magnet, providing a milder magnetically assisted deployment mode.

We next conducted experiments using the capsule on ex vivo porcine stomach tissue to evaluate its performance on a substrate that more closely resembles the native gastric environment (Figure 5E and Movie S8). Under static magnetic guidance, the capsule could still be repositioned on the tissue surface while maintaining its closed configuration, indicating that the hydrogel layer constraint remained stable during transport on a biologically relevant substrate. After AMF-triggered constraint release, the cover recovered through the same underlying mechanism observed on the 3D-printed platform. However, the compliant and adhesive tissue restricted cover displacement, causing it to recover largely in situ rather than propelling the entire capsule away from the deployment site. This localized recovery nevertheless generated sufficient mechanical action to rapidly eject the microrobots together with the PLA base. Viscoelastic deformation and damping of the tissue dissipated part of the stored strain energy. The initial microrobot ejection velocity, estimated from the translational motion of the PLA base, was approximately 0.486 m/s. As the base subsequently separated from the adhesion-constrained cover, the recoil associated with this separation likely imparted an additional impulse, increasing the ejection velocity to 0.852 m/s before subsequent deceleration and landing. After release, the exposed microrobots remained responsive to external magnetic control and could be further guided across the tissue surface. The ex vivo results demonstrate that the capsule can maintain closure, undergo remote on-demand opening, and deploy magnetically controllable microrobots on GI tissue-like biological substrates.

Consistent with the platform experiment, the upright ex vivo test resulted in capsule opening and exposure of the payload chamber without immediate microrobot ejection, and the microrobots were instead drawn out and guided by the external static magnet (Figure S8 and Movie S9).

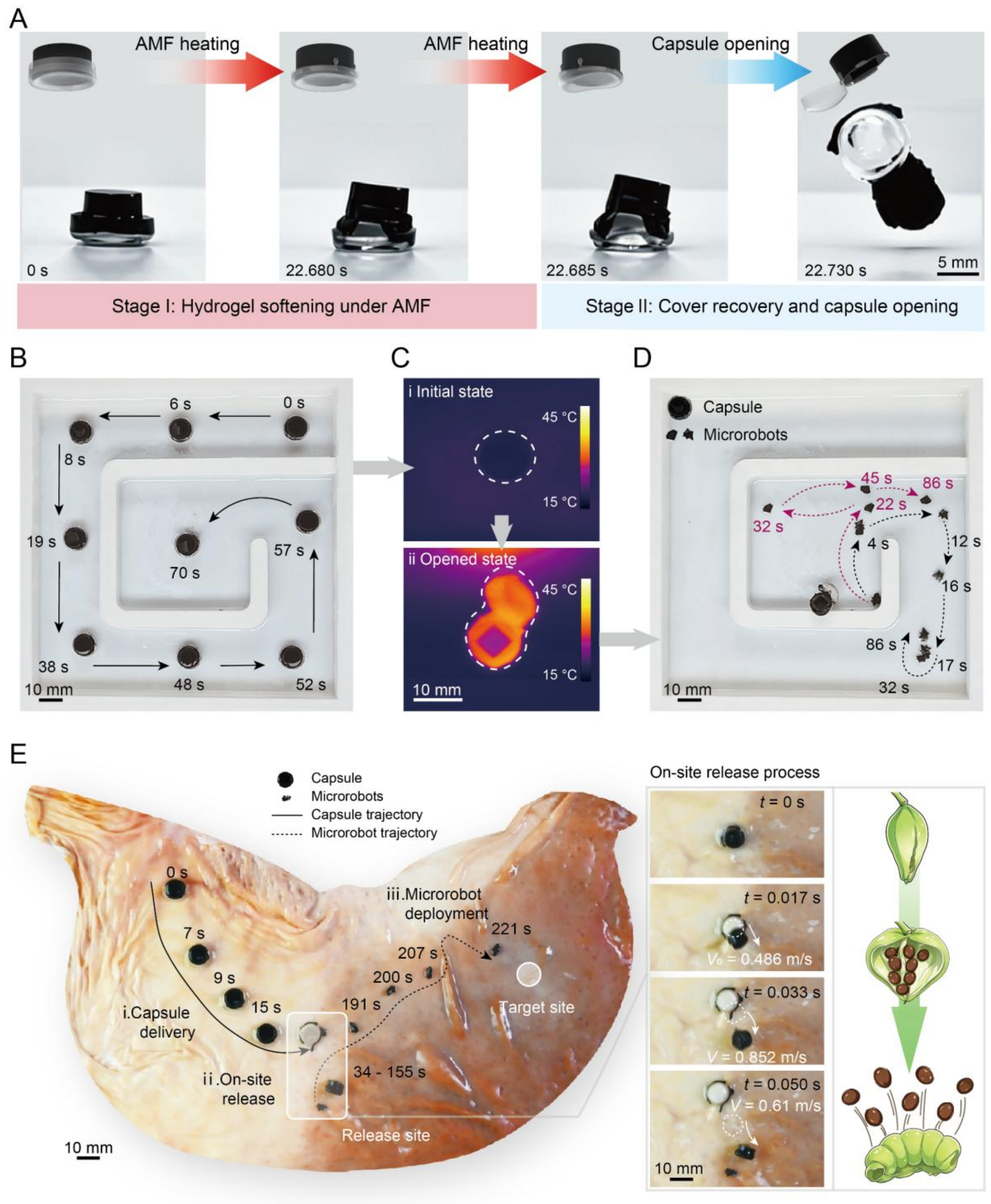

**Figure 5. Functional demonstration of magnetothermally triggered capsule for magnetic delivery and triggered microrobot deployment.** (A) Snapshots of AMF-triggered softening of the hydrogel layer, resulting in cohesive failure within the hydrogel layer and cover recovery

for capsule opening. (B) Magnetic delivery of the capsule to the target location on a 3D-printed platform. (C) Infrared thermal images showing the initial and opened states of the capsule under AMF exposure. (D) Magnetic guidance of released microrobots from the release site to designated position. (E) Magnetic delivery and triggered microrobot deployment by the capsule on ex vivo gastric tissue. The onset of hydrogel fracture and cover recovery was defined as $t$ = 0. The labeled velocities represent the estimated microrobot ejection velocities determined from the translational motion of the PLA base. This release process is analogous to the explosive seed dispersal mechanism of *Impatiens balsamina*.

## 3. Conclusion

In summary, this work presents a magnetothermally triggered capsule that integrates a phase-change hydrogel latch, a monostable cover that serves as an elastic spring, and a PLA base for protected delivery and rapid release of microrobots. Capsule opening results from the hydrogel constraint failure and the release of elastic strain energy stored in the inverted cover. Under the AMF, the hydrogel softens, loses its ability to constrain the cover, and allows the inverted cover to recover spontaneously. The recovered cover converts its stored elastic strain energy into rapid opening and microrobot ejection. Experimental and computational results establish design rules for matching hydrogel rupture resistance with cover restoring energy, preventing premature failure while preserving rapid deployment. System-level demonstrations on a 3D-printed platform and ex vivo porcine stomach tissue show magnetic delivery, target-site opening, microrobot ejection, and post-release magnetic control.

We note that further studies are needed to evaluate long-term safety, in vivo delivery, and deployment under physiological gastrointestinal conditions. The stability of the hydrogel latch under prolonged exposure to gastric and intestinal environments also remains to be systematically investigated, as variations in pH, ionic strength, and fluid exposure may affect its mechanical integrity and triggering behavior. Comparison with reported thermally induced gel-sol hydrogels suggests a trade-off between gel-sol transition temperature and mechanical performance: stronger or more densely crosslinked networks improve mechanical performance but also increase the gel-sol transition temperature (Table S5). Our hydrogel combines a mild gel-sol transition temperature (51.6 °C) with moderate mechanical performance (tensile Young's modulus, 106.1 kPa), although its mechanical performance is limited compared to that of elastomeric cover, which constrains the design of the soft cover. Further improvements in mechanical strength of hydrogel could allow thicker cover designs with greater stored elastic strain energy, thereby enabling higher microrobot ejection speeds. This may be achieved by further developing high-performance, biocompatible magnetothermally responsive materials, although this remains challenging. In addition, the structure of the capsule is a preliminary design, and components such as the base have not yet been fully optimized. Further optimization may enable more precise control of microrobot ejection speed and direction. Despite these limitations, the design principle described here provides a route for matching the restoring energy of the elastic structure with the fracture resistance of latch material to achieve protected delivery followed by rapid release. This strategy may be extended to deployable soft biomedical devices for localized therapy and other applications requiring remote actuation and controlled deployment.

## 4. Experimental Section

*Materials*: Gelatin (gel strength, ~250 g Bloom), Carrageenan (Reagent Grade) and iron oxide nanoparticles ($Fe_3O_4$ NPs, 97% metals basis, 50–300 nm) were supplied by Shanghai Aladdin Biochemical Technology Co., Ltd. The Sylgard 184 polydimethylsiloxane (PDMS) silicone elastomer kit was obtained from Dow Corning. Additional materials are listed in Text S1.1.

*Preparation of hydrogels*: To prepare the magnetothermally triggered phase-change hydrogels, the gelatin solution (20%, w/v) containing different concentrations of carrageenan (0%, 1% and 2%, w/v) was mechanically stirred at 80 °C. $Fe_3O_4$ NPs powder (5%, 10%, and 15%, w/v) was then added, and the mixture was stirred continuously at 60 °C for 30 min. Subsequently, the mixture was poured into molds (consisting of a rubber gasket and two quartz glass plates) and cooled to form hydrogels.

The formulation containing 20% (w/v) gelatin, 1% (w/v) carrageenan, and 15% (w/v) $Fe_3O_4$ NPs was denoted GC-$Fe_3O_4$. Variants were labeled according to the concentration of the component varied from this baseline, as summarized in Table S1.

*Preparation of covers*: Customized molds with different sizes were designed using SolidWorks 2024 (Dassault Systemes, France). Each mold consisted of upper and lower components, which were fabricated by a stereolithography (SLA) 3D printer (Formlabs Form3+, USA). Covers were fabricated by casting PDMS into the corresponding molds. Specifically, the PDMS precursor (base and curing agent mixed at a weight ratio of 10:1) was poured into the lower mold half, after which the upper mold was assembled and secured. The assembled mold was then degassed under vacuum for 20 min to eliminate trapped air bubbles. After curing at 100 °C for 2 h in a drying oven, the covers were carefully demolded from the mold. The preparation process is shown in Figure S9.

*Assembly of the capsule*: The capsule was assembled from three components: a monostable soft cover, a payload-carrying base housing the microrobots, and a magnetothermally triggered phase-change hydrogel layer. First, the base was fabricated by 3D printing from polylactic acid as a cylindrical structure (8 mm in diameter × 5 mm in height). A central square recess (5 mm × 5 mm × 3 mm) was designed in the base to serve as the payload chamber for the microrobots. Next, the hydrogel layer was prepared in bulk and cut into strip-shaped pieces (60 mm × 2 mm × 1 mm). Each hydrogel strip was cut to the required length and then bonded to the upper outer circumference of the base with adhesive, forming a ring-shaped hydrogel constraint above the payload chamber. Finally, the cover was mechanically inverted and then fixed onto the hydrogel layer with adhesive, so that it remained suspended above the base in the inverted configuration, completing the capsule assembly. The exploded view of the assembled capsule is shown in Figure 1D. After capsule assembly, a thin layer of silicone oil was applied to the capsule to reduce water evaporation and improve local heat accumulation of hydrogel during the heating process.

PDMS/NdFeB magnetic elastomer pieces were used as representative magnetic microrobots for capsule loading and post-release magnetic actuation. Detailed fabrication methods are provided in Text S1.2. After the magnetic microrobots were loaded into the payload chamber, the capsule was used in subsequent delivery, triggering, and deployment experiments.

*Characterization of the hydrogels*: The chemical structures and crystalline phases of gelatin, carrageenan, $Fe_3O_4$ NPs and GC-$Fe_3O_4$ hydrogels were characterized by Fourier-transform infrared spectroscopy (FTIR, Vertex 70, Bruker, Germany) and X-ray diffraction (XRD, D8 ADVANCE, Bruker, Germany), respectively.

Rheological measurements were performed using a rotational rheometer (MCR 302, Anton Paar, Austria) with a 20 mm parallel plate. Dynamic temperature sweep tests were conducted to characterize the temperature-dependent hydrogel modulus at 1% strain and 1 Hz frequency. The temperature was ramped from 6 °C to 80 °C at the rate of 4 $°C·min^{-1}$. To minimize water evaporation and reduce heat dissipation during heating, after loading the hydrogel sample beneath the rheometer plate, a thin layer of silicone oil was applied around the plate edge to seal the exposed sample perimeter before the temperature sweep test. Each experiment was repeated at least three times, and representative curves are shown in Figure 2E.

Magnetothermal heating was evaluated using a high-frequency induction heating system operated at a frequency of 50 kHz and an output power of 1 kW, and temperature changes were recorded using infrared thermography. Additional characterization details are provided in Text S1.3.

*Mechanical characterization*: Dumbbell-shaped specimens of hydrogels were prepared by molding according to the ISO 37:2005 Type 3 standard. These dumbbell-shaped samples were tested on a tensile testing machine (Instron 5465, Norwood, USA) equipped with a 1 kN force sensor at a tensile rate of 14 $mm \cdot min^{-1}$ to characterize their mechanical properties.

Peel test: The 180° peel test was conducted to determine the dominant failure mode of the hydrogel layer and quantify its fracture resistance via an energy-balance approach. A previously reported peel mechanics model was adopted to calculate the fracture resistance of the hydrogel.[49] Because the crack propagated within the hydrogel matrix rather than along the adhesive interface, the calculated $G_c$ can represent the cohesive fracture resistance of the hydrogel. Finally, to estimate the initiation stress $\sigma_f$ of the hydrogel, we adopted a standard bilinear cohesive zone relationship. The initiation stress can be derived from the critical energy release rate. Detailed experimental procedures, data analysis, and corresponding equations are available in the Text S1.4.

*Finite element analysis*: To elucidate the nonlinear deformation mechanisms and distinguish the stability of soft covers in the perspective of strain energy, finite element analysis was performed using the commercial software Abaqus. Since the geometry of the capsule could be considered as ideal axisymmetric, a two-dimensional axisymmetric model was constructed and ultimately meshed by 3-node linear axisymmetric elements (CAX3H). The hybrid element formulation was selected to prevent volumetric locking associated with the nearly incompressible nature of the elastomeric matrix. The material of soft cover was modeled using an isotropic, nearly incompressible Mooney-Rivlin hyperelastic model while the neo-Hookean model was adopted to simulate hydrogel. The model parameters were calibrated based on our uniaxial tension test data of PDMS. The parameters used in the software Abaqus were fitted as $C_{10}$ = 0.18 MPa, $C_{01}$ = 0.18 MPa, $D_1$ = 0.05593 $MPa^{-1}$ for the PDMS and $C_{10}$ = 0.0188 MPa, $D_1$ = 1.071 $MPa^{-1}$ for the hydrogel. To capture the limit-point buckling and the snap-through instabilities of soft covers, the modified Riks method was employed. The mechanical loading was simulated by prescribing a rotational displacement to the bottom edge of the bonder structure. To systematically map the stability phase space, an auto submitting process was conducted via a custom Python script. Fracture analysis of hydrogel was based on the comparison between strain energy density along middle path (dangerous section) and the experimentally determined work to rupture $W^*$ (see the detailed analysis steps in Text S1.5).

*Characterization of cover inversion behavior and AMF-triggered capsule opening*: To investigate the effect of cover geometry on monostable and bistable behavior, inversion experiments were conducted using soft covers with different structural parameters. Each cover was manually inverted to its inverted configuration and then evaluated according to its post-inversion response. The manual inversion process and the post-inversion state were recorded using a digital camera (Z5, Nikon). For monostable covers, the inverted configuration was temporarily constrained with tweezers. After the tweezers were released, the spontaneous recovery process was recorded using a high-speed camera (ACS-1, NAC Image Technology) at a frame rate of 20,000 fps. The recorded images and videos were used to determine whether each cover maintained the inverted state, spontaneously recovered to its original configuration, or exhibited intermediate deformation during recovery.

Dynamic opening experiments were performed using assembled capsules. After a thin layer of silicone oil was applied to the capsules, each capsule was placed on a flat platform above a planar helical copper coil that generated the AMF for hydrogel heating. The high-speed camera (ACS-1, NAC Image Technology) was positioned in front of the platform and used to record the opening process at 10,000 fps. The recorded videos were analyzed frame by frame to determine the capsule opening time, cover recovery trajectory, and overall shape evolution during opening.

*Experimental setup for the release and deployment of microrobots*: Magnetically guided delivery and site-specific deployment experiments were conducted on a custom 3D-printed plate. At the beginning of the experiment, the capsule was placed at the initial position on the platform and driven by a permanent magnet along a predefined path toward the target region. After the capsule reached the target site, the permanent magnet was removed, and a planar electromagnetic coil positioned beneath the platform was used to generate AMF, thereby triggering softening of the hydrogel layer and opening of the capsule. The apparatus schematic and magnetic field measurements are provided in Figure S10.

After capsule opening, the permanent magnet was reintroduced to actuate the released microrobots and guide them toward the designated position. The entire process was recorded using a camera (Z5, Nikon), and the heating process was monitored using an infrared thermography camera (PT870, Wuhan Guide Sensmart Technology Co., Ltd., China). Ex vivo validation was performed on porcine stomach tissue to evaluate capsule operation on a more physiologically relevant biological substrate.

**Acknowledgements**

This work was supported by National Natural Science Foundation of China (Nos. 12372167, 12272341, and 12321002), and the Zhejiang Provincial Natural Science Foundation of China under Grant No. LD26A020001, and 111 Project of China (No. B21034).

During the preparation of this work, the authors used ChatGPT (OpenAI) to improve the readability and language of the manuscript. After using this tool, the authors reviewed and edited the content as needed and take full responsibility for the content of the publication.

**Conflict of Interest**

G.M., M.Z., Z.C., and R.X. have filed a patent application covering the magnetothermally triggered capsule technology described in this work. The remaining authors declare no competing interests.

**Author Contributions**

G.M., R.X., M.Z. and Z.C. conceived the project; M.Z. developed the hydrogel and performed the experiments with assistance from Z.C. and Z.L.; Z.C. conducted the numerical simulation and mechanical analysis; M.Z., Z.C., G.M., R.X. and Z.J. analyzed the results; M.Z. and Z.C. wrote the manuscript with comments and materials from all the authors; all authors reviewed and edited the manuscript. G.M. and R.X. supervised the research.

**Data Availability Statement**

The custom Python code used for automated finite element simulations is available from the corresponding authors upon reasonable request. Other data supporting this study are available in this paper and in the supporting information.